\documentstyle[art12]{article}
\date{}
\def\Z{\hbox{{\sf Z}\kern-0.4em {\sf Z}}}

\title{
\Large \bf
\begin{center}
Quantum Group Random Walks in Strongly Correlated $2+1$ $D$ Spin Systems
\end{center}
\magstep 0 { \;
 \\ \\
\vspace{-0.5cm}
\hspace{-1.1cm} \parbox[t]{140mm}{
\begin{center}
A. P. Protogenov, Yu. V. Rostovtsev, and V. A. Verbus,\\
{\footnotesize \em Institute of Applied Physics, Russian Academy of Sciences,
\\
46 Uljanov St., 603600 Nizhny~Novgorod, Russia}
\end{center}
\\ \\
\parbox[t]{140mm}{
We consider the temporal evolution of strong correlated degrees
of freedom in $2+1$~$D$ spin systems using the Wilson
operator eigenvalues as variables. It is shown that the
quantum-group diffusion equation at deformation
parameter $q$ being the $k$-th root of unity has the polynomial solution
of degree $k$.\vspace{0.2cm}\\
{\footnotesize PACS numbers: 74.20.Hi, 74.20.Kk}
}}
\vspace{-1.2cm}
}}

\begin{document}
\baselineskip=20pt

\maketitle

\section*{\small \bf 1. INTRODUCTION}

Strong quantum fluctuations in low dimensional spin systems
change the structure of the ground state and
the classification of low-lying excitations.
The breaking of local magnetic order due to doping and creation of
charge degrees of freedom makes the role of many-body
correlations more essential. For instance, this situation exists
in the layered CuO${}_2$ compounds. The spin disorder of any
number of particles can be expressed in terms of fluxes $\Phi$
of a statistical gauge field
``measuring" a degree of the spin noncollinearity along an
arbitrary large loop $C$.
The statistical magnetic field fluxes through any sets of cells
can be ordered or random depending on the temperature and other factors.
If at low temperature and density the Wigner crystal forms,
then at higher temperature we have the liquid phase of magnetic fluxes.
In the case of Gaussian distribution of fluxes with
the dispersion proportional to the total area $S(C)$,
the mean value satisfies the confinement law
$<\exp{(i\Phi)}> \sim \exp{(-\alpha S(C))}$.
The choice of the gauge and diffeomorphic invariant
eigenvalues $\exp(i\Phi)$ of the Wilson operator
is connected not
only with the need to take into consideration the many-particle correlations.
The choice of these variables simplifies the structure of
the phase space of planar spin systems under consideration.
The finite volume of the phase space
makes the dimension of the Hilbert space finite \cite{Witt89} due to the
periodicity of the canonical conjugated variables $\exp(i\Phi_{a})$
($a=n$ or $a=m$),
for example in the simplest case of the phase space being a torus.
The wave function $\Psi(\exp(i\Phi))$ will be a $k$-dimensional vector,
because the eigenvalue of the Wilson operator is equal to the $k$-th
root of unity, i. e.
$\Phi_n=2 \pi i n /k$ and $n=0,\ldots, k-1$.
Here $k$ is related to the coefficient $K$ before the $SU(2)$
Chern-Simons term in a usual long-wave length description
by $k=K+2$ \cite{Cher89}.
In the general case the quantum number $K$ has a sense of
the linking number of the particle world lines.
In particular, within the $SU_{q}(2)$ quantum group theory of fractional
statistics the
semion excitations have the value $k=K+2=4$, i. e., the $d$-wave
quantum number $K$.

The problems of electronic system properties
in the phase with Gaussian-random or homogeneous distribution of magnetic
fluxes are discussed in a great number of papers published recently
\cite{Pryo92,Alts71,Avis71,Sugi93}. On the other hand, the quantum
group approach to the description of the diffusion in spin chains
has been used in Refs. \cite{Ritt93}.
In this paper we study the $q$-dynamics of the states
characterized by the random distribution of the fluxes with
fixed value of $k$ and an arbitrary value of $n$. We are
interested in the time behavior of transition
processes $n \rightarrow n \pm 1$.

\section*{\small \bf 2. $q$-RANDOM WALKS}

Let us consider the transition from the state
$\Psi(w_n)\!\equiv \Psi_n$
to the neighboring states $\Psi(w_{n\pm 1})$. We use
the parametrization of the wave function argument by the eigenvalue
$w_n = \exp(\Phi_n)$ of Wilson operator
${\hat W} = Tr_{\cal R}\/ \cal P\/
\exp (\oint\limits_{C}{\bf a}\;{\bf dl})$.
Here $\cal R$ denotes an irreducible representation;
$\bf a$ is the non-Abelian gauge potential.
Using these variables the processes of the creation
$\Psi(w_n)\rightarrow \Psi(w_n q)$
and the annihilation
$\Psi(w_n)\rightarrow \Psi(w_n/q)$ of the fluxes
can be considered as hops on the lattice constructed by
the geometrical progression rules.
The denominator of this geometrical progression $q=\exp(2\pi i/k)$
is the $k$-th root of unity. The wave function is now determined
on a discretized circle with the points locating in
the positions of the $k$-th roots
of unity: $\{1, q,\ldots , q^{k-1} \}$.
In this case the procedure
of the comparison of function values in neighboring points is
represented by the difference operator $D_{q}^{+}$:
\begin{equation}
D_{q}^{+}\Psi(w_n) =
\frac{\Psi(qw_n)-\Psi(w_n)}{w_{n}(q-1)}\: ,
\end{equation}
and by Hermite-conjugated ones $D_{q}^{-}$:
\begin{equation}
D_{q}^{-}\Psi(w_n) =
\frac{\Psi(w_{n}\slash q)-\Psi(w_n)}{(q^{-1}-1)}\,w_{n}\: .
\end{equation}
The effect of the Laplace  operator, $\Delta_{q}$ is given by \cite{Koba93}
\begin{equation}
\Delta_{q}\Psi_{n}\equiv
D_{q}^{-}D_{q}^{+}\Psi_{n} =
\frac{\left \Psi_{n+1}+\Psi_{n-1}-2\Psi_{n} \right}
{\left q+q^{-1}-2 \right}
\end{equation}
and the equation of motion in the Euclidean time
\begin{equation}
\frac{\left \partial \Psi\right}{\left \partial t\right}
= \Delta_{q}\Psi
\end{equation}
has the form of the $q$-diffusion equation \cite{Maji93}.
It is easily seen that
the operator (3) coincides with the
Hamiltonian for a rotator
in the limit $k \rightarrow \infty$ $(q \rightarrow 1)$
and Eq. (4) is a usual diffusion equation in this limit.

The eigenfunctions and the eigenvalues of $\Delta_{q}$
are given by
\begin{flushleft}
\begin{equation}
\Psi_{n,\,m}
 = \frac{\left 1 \right}{\left \sqrt k \right}
\,q\,^{mn}\,,
\; \; \; \; \; \;
n,m = 0, 1, \ldots , k-1,
\end{equation}
\end{flushleft}
\begin{equation}
\epsilon_m = \frac{\left \sin^{2}(\pi m\slash k)\right}
{\left \sin^{2}(\pi \slash k)\right}\: .
\end{equation}
By using Eqs. (5), (6) the general solution of Eq. (4) may be
written as
\begin{equation}
\Psi_{n}(t)=\sum\limits_{m=0}^{k-1}a_{m}\Psi_{n,\,m}e^{-\epsilon_{m}t}=
\sum\limits_{m=0}^{k-1}g_{m}G(n-m,\,t)\: ,
\end{equation}
where $g_{n}=\Psi_{n}(0)$ and the Green function $G(n,t)$ has
the following form
\begin{equation}
G(n,\,t)=\frac{1}{k}\sum\limits_{s=0}^{k-1}
\exp \{-\epsilon_{s}t+2\pi isn\slash k\}\: .
\end{equation}

\section*{\small \bf 3. DISCUSSION}

First of all, we would like to draw the attention to the fact that spectrum
(6) has the sense of the squared $q$-dimension $[n]_{q}^2$ of irreducible
representation of algebra $U_{q}(sl_{n})$. Here $
[n]_{q}\equiv {\left(q^{n\slash 2}-q^{-n\slash 2}\right)}\slash
{\left(q^{1\slash 2}-q^{-1\slash 2}\right)}$. From this viewpoint
the temporal evolution looks like
{\em quantum group random walks in the space
of different hooks of the Young tableaux}\/ connected
with representations of the $q$-deformed algebra. This motion
can also be considered \cite{Coqu94} as random
walks on a Dynkin diagram and $t$ as a discrete time parameter.
The statistical models related with random walks on Dynkin
diagrams have been studied in Refs. \cite{PasN87,PasJ87}.
The associated two-dimensional quantum field theories
at critical temperature (equal to $T=1/\ln (1+\sqrt{Q})$,
where $Q=2+q+q^{-1}$ is the Jones algebra index) are conformal field
theories with a value $c=1-6/k(k-1)$ of the central charge.
For $k=2$ we have the usual value $c=-2$ for the random walks.
The connection between the topological properties of random walk
on double punctured plane and the behavior of the four-point
correlation function in the conformal theory with the central
charge $c=-2$ was discussed in Ref.\cite{Nech93}.

It is easily shown using Gaussian-like distribution~(8)
that in the initial state
the mean value $\langle \epsilon_{n}\rangle$, i.e. the variance
\begin{equation}
\langle [n]_{q}^{2} \rangle = \frac{1}{k}\sum\limits_{m=0}^{k-1}
\epsilon_m=\frac{\left 1\right}
{\left 2\sin^{2}(\pi \slash k)\right}\: ,
\label{}
\end{equation}
is proportional to the $q$-analog of the
Casimir operator for odd representations \cite{Wieg94}$.

At $t\to 0$ the variance agrees with the diffusion law
$\langle [n]_{q}^{2} \rangle = {\cal D}\/(k)\, t$
with the diffusion coefficient ${\cal D}\/(k)$,
which equals $1/2$ in the fermion case $k=2$.
For semions ${\cal D}\/(4)=5$.

The $q$-deformed equation of the random walk motion (2) was
first obtained in Ref. \cite{Maji93} (see also
Ref. \cite{MajiPr} in connection with anyon problem).
We  show here that the
Madjid's assumption \cite{Maji93} that there is not reason to think that
the solutions of $q$-diffusion equation
``should exist along familiar Gaussian lines'', is valid
in the following case.

The determination of operator (3) is not unique. Its
determination \cite{Maji93} as $D_{q}^2$ where
$D_{q}\Psi(w) = \{\Psi(qw)-\Psi(w/q)\}\slash\{w(q - q^{-1})\}$
leads to existence of only zeroth modes in the diffusion
equation. In that case Eq.~(4) has the following form:
\begin{eqnarray}
\frac{\partial \Psi}{\partial t}  =
& w^{-2n}[q^{-1}\Psi(w_{n+2})+q\Psi(w_{n-2}) \\
& - (q+q^{-1})\Psi(w_n)] \equiv {\hat A} \Psi. \nonumber
\end{eqnarray}
The multiplier $(q-q^{-1})^{-2}$ in the right-hand side of this
equation is included in the definition of $t$.
The formal solution of Eq.~(10)
\begin{equation}
\Psi_{n}(t) = e^{{\hat A} t}\Psi_{n}(0)
\end{equation}
is a polynomial in $t$ and $q$
\begin{equation}
\Psi_{n}(t) = \sum\limits^{r-1}_{s=0} \frac{t^s}{s !}  {\hat A}^s\Psi_{n}(0)
\end{equation}
\noindent
because the operator $\hat A$ is the nilpotent operator of degree $r$:
${\hat A}^r=0$. The dependence of $r(k)$ is shown in Table 1.\vspace{2.5mm}

\begin{tabular}{|r|r||r|r|r|} \hline
$k$ & odd & even, $4\leq k \leq 10$ & 12 & 14, 16 \\
\hline
$r$ & $(k+1)/2$ & 2 & 3 & 4 \\
\hline
\end{tabular} \vspace{2.5mm}

In the steady-state Eq. (10) is satisfied by the solutions (5)
which are equal to the functions $\Psi_{n}(0)$ in Eq. (12).
To make this fact more clear let us rewrite Eq.~(10)
in the equivalent form
\begin{equation}
\frac{\partial \Psi}{\partial t} = a^2\Psi.
\end{equation}
The annihilation and the creation operators $a$, $a^+$ in the
finite-dimensional $q-$holomorphic representation
are defined by \cite{Flor91}
\begin{equation}
a = g^{-1}\frac{h-h^{-1}}{q-q^{-1}}, \ \ \ \ a^{+}=g,
\end{equation}
where $h = \exp{[(2\pi i/k)P]}$ and $g=\exp{(iQ)}$.
The effect of the last operators is given by
\begin{eqnarray}
(h\Psi) (w_n) = \Psi (w_{n+1}), & \nonumber \\
(g\Psi)(w_n) = w_n\Psi(w_n). &
\end{eqnarray}
They satisfy the following commutator relations
\begin{equation}
h^k = g^k = I, \hspace{3mm} hh^+ = gg^+ = I, \hspace{3mm} hg = qgh.
\end{equation}
Using the latter property it is possible to find \cite{Flor91}
the commutator relation \cite{Bied89,MacF89}
\begin{equation}
aa^+ - qa^+a = q^{-P}.
\end{equation}

The relation shows that
the angular momentum operator P and the particle number operator coincide.
The eigenstates $\Psi_m(w_n)$ of the operator $P$
($P\Psi_m=m\Psi_m, \; m=0, 1, \ldots, k-1$)
coincide with the functions $\Psi_n$ in Eq.~(5).

All eigenvalues, $\varepsilon$ being equal to zero means that
Eq.~(10) describes a motion in the degeneracy space of the ground state.
It is possible, that such a determination corresponds to
the zero value of the $q$-Casimir operator.
In particular, thit takes place
for even representations $U_{q}(sl_{2})$ \cite{Wieg94}.
S.~Nechaev draws our attention to the fact that
the operator in the r.h.s. of Eq. (10)
coincides with the ones for generalized
polynomials $P(L)(l,m)$ for links \cite{Lickor88}.

In connection with the finite Fourier transform (8)
let us pay our attention to
the symmetry of numbers $m$ and $n$ in Eq.(5).
It represents a global gauge invariance of the
choice of polarization in the phase space of canonically conjugated
flux variables $\Phi_{n,m}$.
This invariance is expressed by the invariance with
respect to the action of the modular group,
$SL(2,\,\Z)$ and leads to the fusion of the flux $Z_{n}$-fluctuations.

We would like to formulate the following hypothesis associated with
the dependence of the entropy on the braid linking index $k$.
The operator of the time evolution of states in the frames of
the quantum-group approach can be generalized \cite{Schup93}
with the conservation of energy and nonconservation of entropy,
i.e. with
the possibility of transition from pure states to mixed ones, as
this takes place in open systems. The presence or absence of the
transitions depends on the invariance of the considered mixed
state with respect to the action
of the $q$-generalized evolution operator \cite{Schup93}.
In particular, since the ``renormalized'' coefficient before
the Chern-Simons term for the high-$T_{c}$ superconducting state is zero,
it might be interesting to
check up if the superconducting coherent state differs in the
above respect from the high-temperature "normal" state which has
partially conserved the quantum-group order.

In conclusion, we have performed a study of
the $q$-Brownian motion in the flux space.
For the deformation parameter being a root of unity, it is found out
that the general solution of the $q$-diffusion
equation is the $q$-power polynomial and the distribution
of the flux numbers is characterized by the variance that is proportional
in the initial state to the $q$-analog of the Casimir operator.

\section*{\small \bf ACKNOWLEDGEMENTS}

This work was supported in part by a Soros Foundation Grant
awarded by the American Physical Society.
We would like to thank M. A. Antonets,
V. A. Mironov, V. E. Semenov, and I. A. Shereshevsky
for many helpful and stimulating discussions.
It is a pleasure to acknowledge the support of A. G. Litvak.
One of the authors (A.P.P.) would like to thank Professor
Abdus Salam, the International Atomic Energy Agency
and UNESCO for hospitality at the the International Centre
for Theoretical Physics, Trieste.

\end{document}